\documentclass[12pt,twoside]{article}   
\usepackage[super,sort&compress,comma]{natbib}

\usepackage{fancyhdr}		

\usepackage[section]{placeins}   %

\usepackage{url}
\usepackage{caption} 
\usepackage{framed,multirow}

\usepackage{amssymb}
\usepackage{amsmath}
\usepackage{latexsym}

\usepackage{color, soul}
\usepackage{xargs}
\usepackage{tikz}
\usepackage{todonotes}
\usepackage{multirow} 
\usepackage{verbatim}
\usepackage{nameref,hyperref}

\usepackage{multicol}
\usepackage{multirow}
\usepackage{graphicx}
\usepackage{xcolor,colortbl,xspace}
\usepackage{comment}
\usepackage{chngcntr}
\usepackage{url}
\usepackage{xcolor}

\usepackage{hyperref}
\usepackage{subcaption}
\usepackage{bm}

\usepackage{graphicx}

\makeatletter \renewcommand\@biblabel[1]{$^{#1}$} \makeatother
\usepackage{booktabs}

\usepackage{hyperref}
\hypersetup{ colorlinks,
    citecolor=blue,
    filecolor=blue,
    linkcolor=blue,
    urlcolor=blue
}

\usepackage{xcolor}

\definecolor{gray}{rgb}{0.6,0.6,0.6}
\definecolor{red}{rgb}{0.85,0,0}
\definecolor{green}{rgb}{0,0.85,0}
\definecolor{blue}{rgb}{0,0,0.85}
\definecolor{beige}{rgb}{0.92,0.87,0.78}
\usepackage[all]{hypcap}    

\newlength\savewidth\newcommand\shline{\noalign{\global\savewidth\arrayrulewidth
  \global\arrayrulewidth 1.5pt}\hline\noalign{\global\arrayrulewidth\savewidth}}

\begin{document}
\title{Self-supervised learning improves robustness of deep learning lung tumor segmentation to CT imaging differences}
\author{Jue Jiang, Aneesh Rangnekar, Harini Veeraraghavan
\thanks{The manuscript has been submitted on XXX, 2024 for review. This work was partially supported by the NCI R01CA258821 and the MSK Cancer Center core grant P30 CA008748. Jue Jiang and Harini Veeraraghavan contributed equally. All authors are with the Department of Medical Physics, Memorial Sloan Kettering Cancer Center, NY, New York, USA.(e-mail: jiangj1@mskcc.org and veerarah@mskcc.org)}}
\maketitle

\begin{abstract}
\textbf{Background:} Self-supervised learning (SSL) is an approach to extract useful feature representations from unlabeled data, and enable fine-tuning on downstream tasks with limited labeled examples. Self-pretraining is a SSL approach that uses the curated task dataset for both pretraining the networks and fine-tuning them. Availability of large, diverse, and uncurated public medical image sets provides the opportunity to apply SSL in the "wild" and potentially extract features robust to imaging variations. However, the benefit of wild- vs self-pretraining has not been studied for medical image analysis.
\\
\textbf{Purpose:} Compare robustness of wild versus self-pretrained transformer (vision transformer [ViT] and hierarchical shifted window [Swin]) models to computed tomography (CT) imaging differences for non-small cell lung cancer (NSCLC) segmentation. 
\\
\textbf{Methods:} Transformer models (ViT and Swin) were pretrained with SSL wild- approach with 10,000 unlabeled and uncurated CTs sourced from The Cancer Imaging Archive and internal datasets. Self-pretraining was applied to same architectures using a curated public task dataset (n = 377) of patients with NSCLC. Both pretraining approaches were performed using pretext tasks introduced in self-distilled masked image transformer approach. All models were fine-tuned to segment NSCLC (n = 377 training dataset) and tested on two separate datasets containing early (public N = 156) and late stage (internal N = 196) NSCLC. Models were evaluated in terms of: (a) accuracy, (b) robustness to image differences from contrast, slice thickness, and reconstruction kernels, and (c) impact of pretext tasks for pretraining. Feature reuse was evaluated using centered kernel alignment. 
\\
\textbf{Results:} Wild-pretrained Swin models outperformed self-pretrained Swin for the various imaging acquisitions. ViT resulted in similar accuracy for both wild- and self-pretrained models. Masked image prediction pretext task that forces networks to learn the local structure resulted in higher accuracy compared to contrastive task that models global image information. Wild-pretrained models resulted in higher feature reuse at the lower level layers and feature differentiation close to output layer after fine-tuning.
\\
\textbf{Conclusion:} Wild-pretrained networks were more robust to analyzed CT imaging differences for lung tumor segmentation than self-pretrained methods. Swin architecture benefited from such pretraining more than ViT. 
\end{abstract}

\maketitle


\section{Introduction}
Self-supervised learning (SSL) has emerged as a promising approach to train large deep learning (DL) models in both computer vision and medical image analysis applications, especially when labeled data for various tasks are limited. This is because, SSL can be used to pretrain the model with unlabeled instances, following which it can be fine-tuned with relatively few labeled examples from downstream task datasets. Pretraining with unlabeled instances with SSL is made possible by the use of pretext tasks, whereby supervised tasks are created to make use of the unlabeled data as `ground truth'. Examples of such pretext tasks include prediction of masked image portions\cite{he2021masked, jiang2022self_SMIT}, masked image reconstruction\cite{tang2022self}, jigsaw puzzles\cite{zhou2021models,zhu2020rubik}, as well as combination of contrastive and image reconstruction losses\cite{taleb20203d,zhou2023unified_PRCLV2}.

Self-pretraining is a commonly used SSL approach for medical images, wherein the task-specific curated dataset is used to pretrain the model without exposing the labels, and then subsequently used with the labels to fine-tune the model using supervised training\cite{zhou2021models,chen2021transunet,Umapathy2023MedPhys}. SSL pretraining in the `wild' is a different approach, wherein large and diverse datasets that are uncurated and unrelated to task are used to pretrain the model without labels. Wild-pretraining with massive unlabeled datasets forms the basis of training high-capacity foundation models\cite{cheng2023sam,liu2023clip}, and has demonstrated effectiveness for natural image analysis\cite{Matsoukas2022}. Open-source datasets from the Cancer Imaging Archive (TCIA)\cite{clark2013cancer} have been used to successfully pretrain medical image models and then successfully applied to multi-organ segmentation following fine-tuning with relatively few examples\cite{tang2022self,jiang2022self_SMIT,Qayyum2023BHI,NguyenAAAI2023,YanWACV2023}. A key difference between self- and wild-pretraining is that the former approach uses curated task-specific datasets to learn feature representations as opposed to the latter that seeks to extract feature representations from uncurated and large datasets. 

The premise for using wild-pretraining is that models trained with images with substantial imaging variations would extract universally applicable feature representations that lead to robustly accurate models for clinical and multi-institutional research studies. However, these relative merits of wild-pretraining over self-pretraining in terms of robustness to imaging acquisition differences has not been studied, particularly for tumor segmentation. Tumor segmentation is interesting in this context because uncurated datasets may not contain sufficient or even any representation of the studied tumor in the pretraining dataset as opposed to task-specific datasets used for self-pretraining that contain the specific tumor types. 

In addition to the training strategy, the choice of network architectures can also influence accuracy due to the differences in the inductive bias contained in the network. For instance, vision transformer (ViT), which uses a sequence of image tokens obtained by tokenization of non-overlapping input image patches, has the lowest inductive bias as it eliminates all relative spatial and local relationships. Hierarchical shifted window transformer (Swin) adds back some of the inductive bias by employing cross-window attention and hierarchical scaling. Convolutional neural network (CNN) has the highest inductive bias as it maintains the spatial locality of the images along the different feature layers. Hence, we analyzed ViT, Swin, and a CNN-based non-skip U-Net to assess the potential benefits of the two pretraining strategies for these architectures.

Finally, we studied why accuracy differences occur by performing feature reuse analysis using centered kernel alignment (CKA). CKA provides insights into what extent the features in the different layers were modified as result of fine-tuning. Higher feature reuse in the lower level layers has shown to impart improved accuracy on downstream tasks on computer vision\cite{Goyal2021} and medical image analysis tasks\cite{Matsoukas2022}. 

Our contributions are: (a) Comparative analysis of different SSL-based wild-pretraining and self-pretraining applied to three common architectures, a ViT, Swin, and a U-Net-based CNN for lung tumor segmentation, (b) analysis of robustness to CT acquisition differences due to the two SSL pretraining for the individual architectures; we evaluated the performance of the various models on an open-source CT lung tumor phantom dataset to assess their capability to segment images that are different from real patient scans. And (c) we studied how feature reuse was influenced by SSL pretext tasks as well as the choice of network and its impact on downstream accuracy. 

To our best knowledge, this is the first study to evaluate the relative benefits of wild vs. self-pretraining and further our understanding of how and why segmentation robustness to CT imaging variations occurs from such methods. Understanding the relative merits of the pretraining methods could inform the development of pretrained models in clinical and research use as well as understand the implications of such models when deployed in clinical practice. 

\section{Methods}
\subsection{Background terminology}
\begin{itemize}
    \item \textbf{Self-supervised learning (SSL): \/}\rm An unsupervised machine learning approach that creates pretext tasks that use the unlabeled data as the "ground truth" to extract useful features that are then applicable to downstream tasks. 
    \item \textbf{Self-pretraining: \/}\rm A SSL approach that applies unsupervised pretraining on the same curated task-specific dataset that is to be used for supervised fine-tuning.
    \item \textbf{Wild-pretraining: \/}\rm An unsupervised pretraining approach applied with large and relatively uncurated datasets that do not share the same characteristics (disease, disease site, imaging characteristics) as the curated task-specific downstream dataset. 
    \item \textbf{Upstream versus downstream dataset: \/}\rm Pretraining dataset is used to pretrain network with pretext tasks. A downstream dataset is a labeled dataset used for fine-tuning the network for performing specific tasks. 
    \item \textbf{Pretext task: \/}\rm Tasks created for optimizing the network using supervised losses using the unlabeled data as ground truth, e.g. predicting masked image portions, predicting rotation of rotated input images, reconstructing input image corrupted by maksing, etc. 
\end{itemize}

\begin{table}[t]
\vspace*{1ex}
\centering
\caption{Summary of wild- pretraining, self-pretraining, fine-turning, and testing datasets. $\dagger$: used in self-pretraining and subset of 316 CTs was included in standard pretraining. Recon 1 kernels: GE ''standard'' and ''bone'', Siemens with $<$ B40; Recon 2: GE ''Bone Plus'', Siemens with $\geq$ B40 and $<$ B50; Recon 3: GE ''Lung'', Siemens $\geq$ B50.}
\label{tab:dataset}
\def\arraystretch{1.25}
    \resizebox{1.0\textwidth}{!}{
    \large
    \begin{tabular}{*{7}{l}}
       Data & Location & Number &  Manufacturer & Thickness & Kernels & Contrast Information \\
        \shline
        \textbf{Wild-pretraining} & & & & & & \\
        MELA 2022\cite{xiao2023lesion} & Chest & 880 & NA & 1 mm & NA & contrast, non-contrast\\
        AMOS 2022\cite{ji2022amos} & Chest-Abd-pelvis & 360  & NA & 5 mm to 7.5 mm & NA & contrast, non-contrast \\
        COVID-19\cite{harmon2020artificial} & Chest & 609 & NA & 5 mm & NA & contrast, non-contrast \\ 
        KITS \cite{C4KC-KiTS} & Abdomen-pelvis & 411 & \begin{tabular}[c]{@{}l@{}}Siemens, \\ Toshiba\end{tabular} & 3 mm & smooth & \begin{tabular}[c]{@{}l@{}} arterial, late, \\ non-contrast\end{tabular} \\

        Pancreas CT\cite{roth2015deeporgan} & Chest-Abdomen & 80 & NA & 1 mm to 5mm & NA & contrast, non-contrast  \\ 
        Internal 1 Radiotherapy & Chest & 5,124 & GE & 3 mm to 5 mm & smooth, sharp & contrast, non-contrast \\
        Internal 2 Radiotherapy & Head and neck & 2,632 & GE & 2.5 mm to 3 mm & smooth & contrast, non-contrast\\
         \midrule
         \textbf{Fine-tuning$^\dagger$} & & & & & & \\
         TCIA NSCLC\cite{aerts2015data} & Chest-abdomen & 350  & Siemens, CMS & 3 mm & Smooth & contrast, non-contrast \\ 
         TCIA NSCLC\cite{aerts2015data} & Chest-abdomen & 279  & Siemens & 3 mm & Smooth & non-contrast \\ 
         \midrule
         \textbf{Testing} & & & & & & \\
         LRad\cite{bakr2018radiogenomic} & Chest & 156 & \begin{tabular}[c]{@{}l@{}}Siemens, \\ Toshiba, GE\end{tabular} & 0.9 mm to 5 mm & \begin{tabular}[c]{@{}l@{}}smooth, medium, \\ sharp \end{tabular} & contrast, non-contrast \\
         LC & Chest-abdomen & 196& GE & 1.25 mm to 5mm & smooth, sharp & contrast, non-contrast \\ 
         Lung Phantom \cite{zhao2015} & Chest & 1 & GE & 1.25 mm & smooth & contrast \\
       \bottomrule
    \end{tabular}
    }
\vspace*{1ex}
\end{table}

\subsection{Datasets}
Analyzed datasets with imaging acquisitions, and disease details are in Table~\ref{tab:dataset}. Retrospective analysis of institutional datasets were approved with local institutional review board, with a waiver for written informed consent, and was compliant with the Health and Insurance Portability and Accountability Act. Due to the wide variation in the convolution kernels used in the testing datasets, we categorized them as Recon 1 kernels: GE ''standard'' and ''bone'', Siemens with $<$ B40; Recon 2: GE ''Bone Plus'', Siemens with $\geq$ B40 and $<$ B50; Recon 3: GE ''Lung'', Siemens $\geq$ B50. 

\textbf{Wild-pretraining: \/}\rm A total of 10,412 3D CT scans encompassing diseases from head and neck to the pelvis sourced from datasets provided publicly for variety of tasks including lesion detection~\cite{xiao2023lesion}, classification~\cite{harmon2020artificial}, and multi-organ and abdominal tumor segmentations were used without additional curation for pretraining. Anonymized institutional datasets were used as is without any curation and consisted of CT scans from patients treated for lung, esophageal (Internal 1) and head and neck (Internal 2) cancers treated with radiotherapy (RT). 

\textbf{Fine-tuning and self-pretraining: \/}\rm The public domain fine-tuning dataset consisted of patients diagnosed with locally advanced non-small cell lung cancer (LA-NSCLC) who underwent RT in a single institution\cite{aerts2015data}. Tumor contours are provided with the dataset. Patients had contrast or non-contrast CTs and typically reconstructed with convolution kernels ($\leq$ B30). Tumor sizes ranged with a median of 33.68 cc, and interquartile range (IQR) of 8.29 cc to 90.31 cc. A random set of 316 cases were included in unsupervised pretraining and self-pretraining.

\textbf{Testing: \/}\rm Two held-out testing datasets were analyzed: (1) a public dataset (LRad) containing early stage (stage I-II) lung cancer, all of whom underwent surgery\cite{bakr2018radiogenomic}, (2)  a dataset with primary and metastatic stage III-IV lung lesions (LC). Tumors in LRad ranged in size with a median of 7.91cc, IQR of 3.60 cc to 28.23 cc, whereas the tumors in LC ranged in size with a median of 19.54 cc, IQR of 6.64 cc to 66.97 cc. A subset of 20 cases in LC had patients reconstructed with 2.5mm, 5mm slice thickness as well as with GE's lung kernel and GE's standard kernel and were evaluated to assess pairwise accuracy differences. Finally, a single lung CT phantom dataset with 8 different lesions of varying shapes and sizes provided through the TCIA repository\cite{zhao2015} was also evaluated. The lung phantom image scan consisted of 8 lesions of different shape and sizes, scanned at Columbia University Medical Center with GE scanner at 120 kVp. The images were reconstructed using a lung kernel with 1.25 mm slice thickness.

\begin{figure}[t]
\vspace*{1ex}

    \centering
    \begin{subfigure}[b]{0.48\textwidth}
        \centering
        \includegraphics[width=\textwidth]{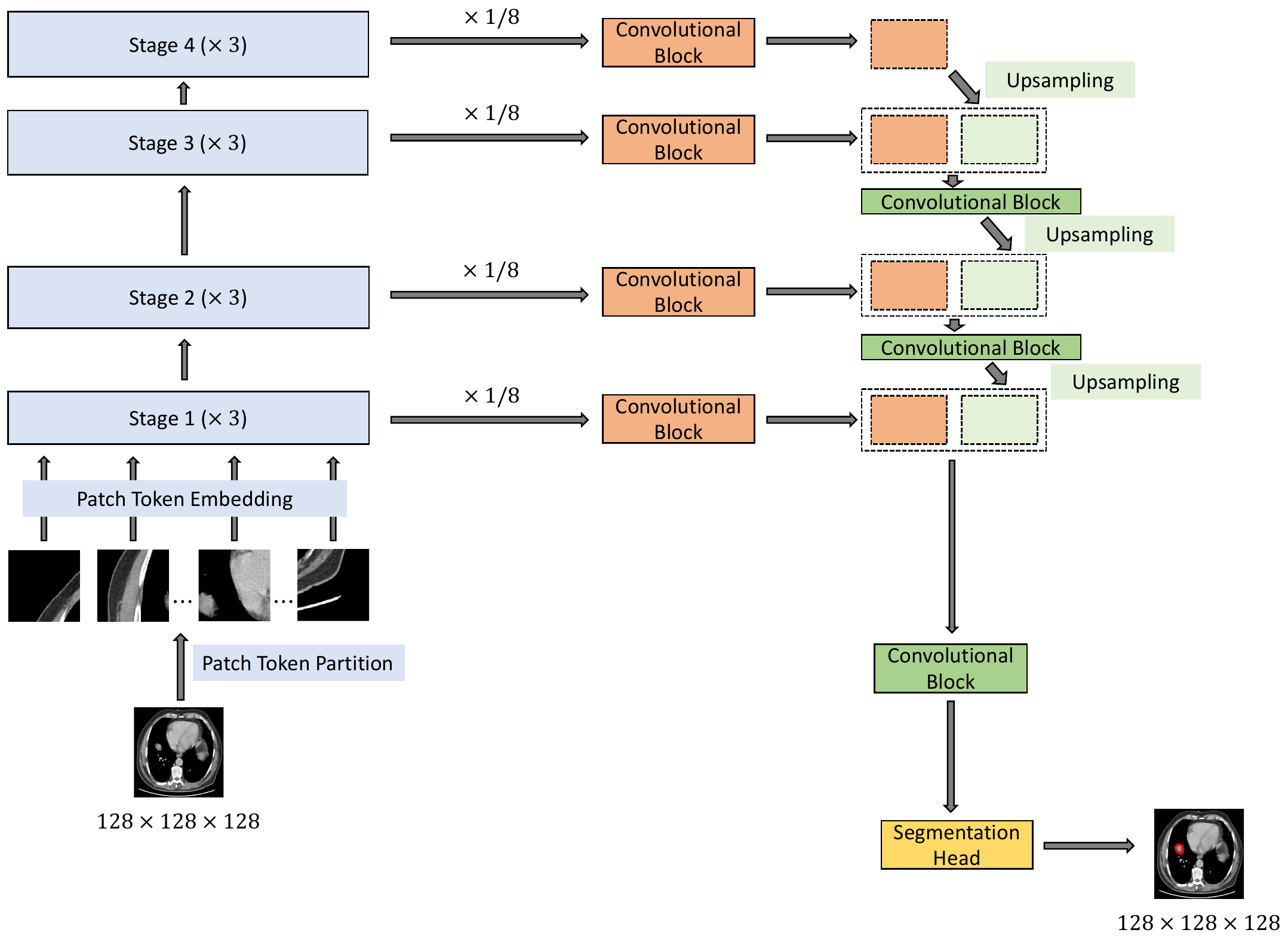}
        \caption{ViT (Vision Transformer)}
        \label{fig:sub2}
    \end{subfigure}
    \hfill
    \begin{subfigure}[b]{0.48\textwidth}
        \centering
        \includegraphics[width=\textwidth]{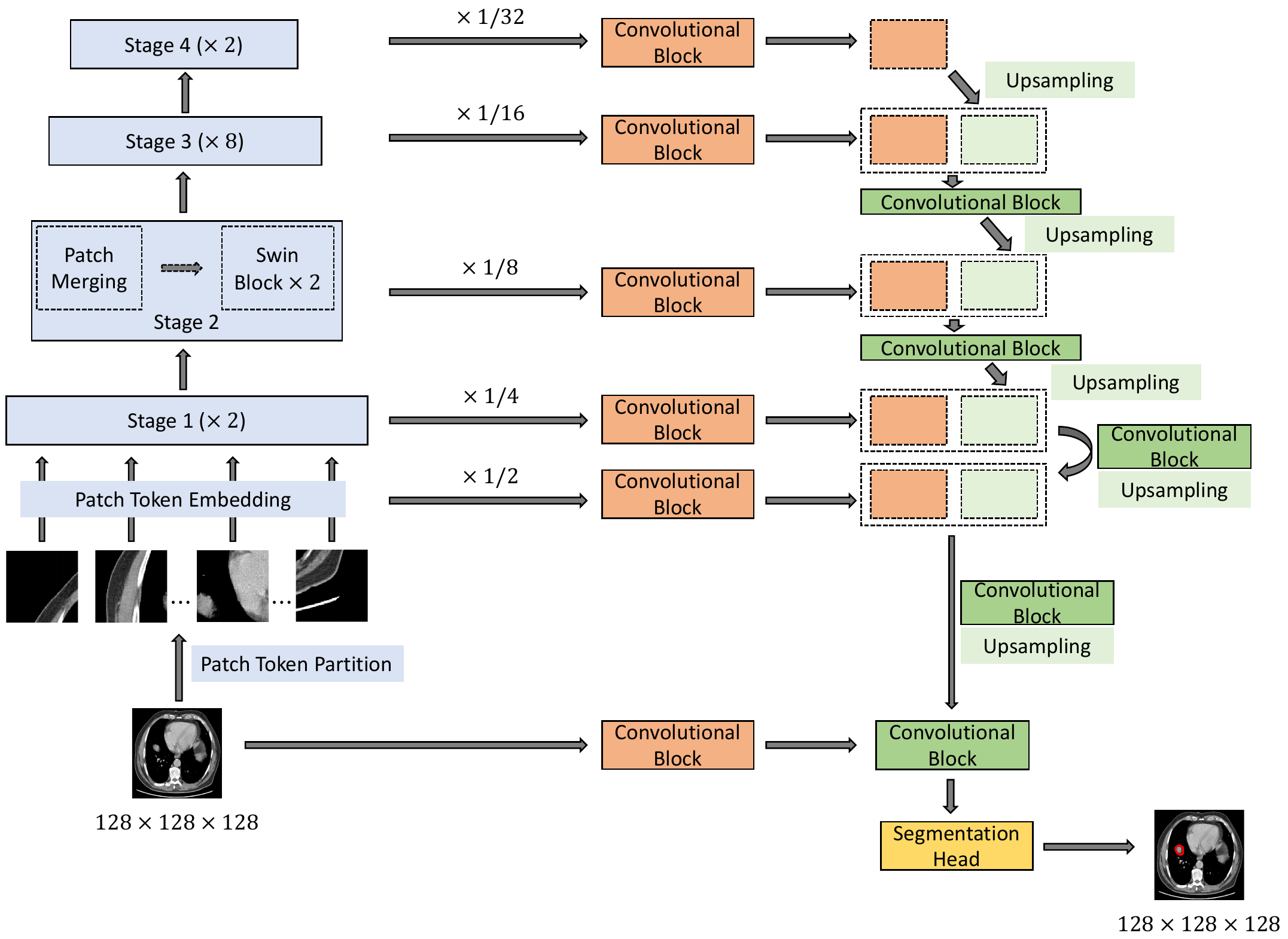}
        \caption{Swin (Hierarchical Transformer)}
        \label{fig:sub1}
    \end{subfigure}
    \caption{Transformers architectures combined with the convolutional decoders for 3D tumor segmentation.}
    \label{fig:subfigures}
\vspace*{1ex}
\end{figure}

\subsection{Network architectures}

We selected two commonly used 3D transformer encoder architectures in medical image segmentation tasks\cite{hatamizadeh2022unetr,tang2022self,chen2023masked,chen2021transunet}, namely the Vision transformer\cite{dosovitskiy2021} (ViT) and hierarchical Swin Transformer\cite{liu2021swin} (Swin). Figure~\ref{fig:sub2} and  Figure~\ref{fig:sub1} show the corresponding schematics of the 3D ViT and 3D Swin networks that use transformer blocks in the encoder and convolutional blocks with skip connections in the decoder path. The ViT encoder comprised of 12 transformer blocks, 8 multi-head self attention at each stage, and a feature embedding size of 768. The ViT network operated on a patch size of 8 $\times$ 8 $\times$ 8. The 3D Swin network used the Swin encoder backbone, with a depth of [2,2,8,2] and corresponding [4,4,8,16] multi-head attention for the four stages of hierarchical transformer blocks. It operated on a patch size of 2 $\times$ 2 $\times$ 2, a window size of 4 $\times$ 4 $\times$ 4, and the feature embedding size of 384. An additional convolutional block was added to the decoder to process the extracted encoder feature in both ViT and Swin architectures. The total number of parameters for ViT and Swin-based segmentation models were 46,405,874 and 64,698,114, respectively.\\ 
In addition, a non-skip U-Net as used in a prior work\cite{zhou2023unified_PRCLV2} was used as a baseline to assess the relative merits of self- and wild-pretraining on a network using only convolutional layers, \textcolor{black}{which has 17,111,499 parameters}. All networks used an image size of 128 $\times$ 128 $\times$ 128.

\section{SSL-pretext tasks for unsupervised pretraining}
Pretext tasks in SSL are used to train models to `learn' the underlying feature representations within the images by providing supervisory signals without relying on external user-annotated labels. Hence, pretext tasks are designed to use the images themselves as labeled instances to perform supervised network optimization. We used masked image prediction (MIP), wherein random 3D patches of the input image are masked out and the network must predict the missing image patches by using the context of visible patches. We also used masked patch token distillation (MPD) and image token distillation (ITD) tasks that were introduced in a prior work using a self-distillation approach to perform SSL\cite{jiang2022self_SMIT}. Additionally, we also investigated a contrastive learning\cite{taleb20203d} SSL task for both wild and self pretraining. 

Self-distillation learning was performed by creating a `teacher' network with identical architecture as the student by using exponential moving average of the `student' model. The student model is subjected to supervised optimization objectives and is presented with masked images. The teacher on the other hand is presented with a clean and different 3D cropped view of the original image. 

MIP involves a dense pixel regression of the pixel intensities within the masked image patches and is optimized by minimizing the L1 norm between the predicted image patches and the corresponding unmasked patches. The loss for optimizing MPD task was implemented using cross-entropy loss with temperature scaling to adjust the sharpness of the token distribution produced by the student and teacher network. Finally, ITD matched the global class embedding computed by student and teacher networks by minimizing the cross-entropy loss. 

Furthermore, the pretext task, which combines contrastive and multi-scale image reconstruction tasks and is referred to as PCRLv2\cite{zhou2023unified_PRCLV2}, was also analyzed using a CNN backbone.

\subsection{Fine-tuning details}
All six networks (3 self- and 3 wild-pretrained) for ViT, Swin and CNN as well as networks pretrained with different pretext tasks as part of ablation analysis were fine-tuned on an identical downstream dataset to generate volumetric segmentation of lung tumors. The segmentation model was optimized using a combination of cross-entropy and Dice overlap losses.  

\subsection{Implementation details}
Pretraining (self and wild) was performed by generating two augmented views, one masked view provided to student using a default masking ratio of 0.75 and the second unmasked view provided to the teacher, respectively. All images were clipped in the HU value of [-500,500], normalized to [0,1] and resampled to a uniform voxel spacing of 2mm $\times$ 2mm $\times$ 2mm and then randomly cropped to 128 $\times$ 128 $\times$ 128 voxels to generate the 3D views. The networks were optimized using ADAMw~\cite{Loshchilov2017DecoupledWD} with a cosine learning rate scheduler~\cite{Loshchilov2016SGDRSG} trained for 500 epochs with an initial learning rate of $8e^{-4}$ and warmup for 50 epochs. In order to account for fewer unlabeled examples available for self-pretraining, the self-pretraining was performed for 2,000 epochs with a warmup for 200 epochs. Online data augmentations were performed to increase data variations. A path drop rate of 0.1 was applied to the student model, and all experiments were conducted on 4 NVIDIA A100 GPUs (4 $\times$ 80GB memory) using an effective batch size of 8 for ViT and 32 for Swin transformers.
\\
Fine-tuning used only the student network and was performed on 4 NVIDIA A100 GPUs. All analyzed network configurations were trained with a learning rate of $2e^{-4}$ for 1,000 epochs. The Swin models were fine-tuned with a batch size of 24 while the ViT models used a batch size of 4 due to memory limitations. Fine-tuning and testing was performed after resampling all the images to a uniform voxel spacing of 1.5 mm $\times$ 1.5 mm $\times$ 2.0 mm voxels. Early stopping selected the model with the best validation accuracy to prevent over-fitting. 

\subsection{Metrics and statistical analysis}
Tumor detection rate (DR) was computed for all of the analyzed methods and subjected to self- or wild-pretraining using the various pretext tasks. DR is defined as the overlap between the segmentation and ground truth was at least $\tau=0.5$ \cite{jiangTMI2019}. Tumor segmentations were compared against manual delineations using Dice similarity coefficient (DSC) and Hausdroff distance at 95$^{th}$ percentile (HD95). Segmentation differences between self-pretraining and wild-pretrained models were measured using paired, two-sided, Wilcoxon signed rank tests at 95\% significance level. 
\\
Centered kernel alignment (CKA) was used to measure feature reuse between the wild-pretrained and self-pretrained models before and after fine-tuning. CKA computes a normalized similarity of two feature representations $\bm{M}$ and $\bm{N}$ in terms of the Hilbert-Schmidt Independence Criterion (HSIC):\\

\begin{equation}
\setlength{\abovedisplayskip}{1pt}
\setlength{\belowdisplayskip}{1pt} 
 \mathrm{CKA}(\bm{X},\bm{Y}) = \frac{\mathrm{HSIC_0}(\bm{X},\bm{Y})}{\sqrt{\mathrm{HSIC_0}(\bm{X},\bm{X})\mathrm{HSIC_0}(\bm{Y},\bm{Y})}}
\label{eqn:CKA}
\end{equation}  
where $\bm{X}$=$\bm{M} \bm{M^T}$ and $\bm{Y}$=$\bm{N} \bm{N^T}$ are the Gram matrices of feature $\bm{M}$ and $\bm{N}$ respectively. CKA computation requires the feature activations of entire dataset to be stored in the memory, which is difficult to implement for transformers that have a large number of parameters. Minibatch CKA was computed by averaging HSIC scores over k minibatches \cite{nguyen2020wide} as:
\begin{equation}
\setlength{\abovedisplayskip}{1pt}
\setlength{\belowdisplayskip}{1pt} 
 \mathrm{CKA_{minibatch}} = \frac{\mathrm{\frac{1}{k}\sum_{i=1}^k HSIC_1}(\bm{\mathrm{M_iM_i^T}},\bm{\mathrm{N_iN_i^T}})}{\sqrt{\mathrm{\frac{1}{k}\sum_{i=1}^k HSIC_1}(\bm{\mathrm{M_iM_i^T}},\bm{\mathrm{M_iM_i^T}})} \sqrt{\mathrm{\frac{1}{k}\sum_{i=1}^k HSIC_1}(\bm{\mathrm{N_iN_i^T}},\bm{\mathrm{N_iN_i^T}})} }
\label{eqn:CKA_minibatch_supp}
\end{equation}  
where $M_i$ and $N_i$ are matrices containing activations of the $i^{th}$ minibatch. An unbiased estimator of HSIC\cite{song2012feature} was computed to mitigate dependency of CKA values on the batch size:
\begin{equation}
\setlength{\abovedisplayskip}{1pt}
\setlength{\belowdisplayskip}{1pt} 
 \mathrm{HSIC_1}(\bm{X},\bm{Y}) = \frac{1}{n(n-3)} (  \mathrm{tr}(\tilde{X}\tilde{Y}) + \frac{\bm{1}^\mathrm{\bm{T}}\tilde{X}\bm{1}\bm{1}^\mathrm{\bm{T}} \tilde{Y} \bm{1}}{(n-1)(n-2)} - \frac{2}{(n-1)}\bm{1}^\mathrm{\bm{T}} \tilde{X}\tilde{Y}\bm{1})
\label{eqn:CKA_minibatch_bias_est}
\end{equation} 
where $\tilde{X}$ and $\tilde{Y}$ were obtained by setting the diagonal entries of the similarity matrices $X$ and $Y$ to zero.

\section{Experiments configuration}
In all experiments, the performance comparisons were done between self- and wild-pretrained model instances of the same architectures. Robustness to CT acquisition differences was measured by comparing the segmentation accuracies for Recon 1 versus Recon 2 versus Recon 3 kernels, contrast and non-contrast CTs, and different slice reconstruction thicknesses. In addition, the impact of pretext losses on downstream accuracy as well as feature reuse was evaluated for both self- and wild-pretraining configurations for the Swin network. We evaluated whether sequencing wild- and self-pretraining improves tumor segmentation accuracy. Finally, we also studied the impact of fine-tuning the two transformer networks with only non-contrast CTs from a single scanner manufacturer, namely Siemens, on the downstream accuracy. 
 
\section{Results}

\begin{table}[t]
\vspace*{1ex}
\centering
\caption{Tumor detection and segmentation accuracy with pretraining methods and transformer architectures. DR: Detection rate.}
\label{tab:tumor_seg_acc}
\def\arraystretch{1.25}
\resizebox{0.85\textwidth}{!}{%
\large
\begin{tabular}{lll|lll|lll}
\multirow{2}{*}{Model} & \multirow{2}{*}{Training Strategy}         & \multirow{2}{*}{\begin{tabular}[c]{@{}l@{}}Pretext \\ Task\end{tabular}} & LRad      &  &  & LC  &             &         \\
      &                  &              & DSC                    & HD$_{95}$               & DR     & DSC                   & HD$_{95}$                 & DR   \\
\shline
CNN   & Scratch          & N/A          & 0.42$\pm$0.34          & 11.99$\pm$8.34          & 0.48      & 0.54$\pm$0.27          & 12.36$\pm$10.37          & 0.60    \\
CNN   & Self-pretraining & PCRLv2       & 0.45$\pm$0.33          & \textbf{10.07$\pm$8.32} & 0.52      & \textbf{0.56$\pm$0.23} & \textbf{12.50$\pm$10.93} & 0.67    \\
CNN   & Wild-pretraining & PCRLv2       & \textbf{0.46$\pm$0.33} & 11.99$\pm$8.34          & 0.54      & 0.54$\pm$0.24          & 12.81$\pm$10.97          & 0.62    \\
\midrule
ViT   & Scratch          & N/A          & 0.55$\pm$0.31          & 8.71$\pm$7.51           & 0.60      & 0.68$\pm$0.27          & 12.76$\pm$20.51          & 0.74    \\
ViT   & Self-pretraining & SMIT         & 0.64$\pm$0.25          & 8.59$\pm$7.53           & 0.65      & 0.71$\pm$0.22          & \textbf{10.97$\pm$10.62} & 0.80    \\
ViT   & Wild-pretraining & SMIT         & \textbf{0.66$\pm$0.22} & \textbf{8.52$\pm$7.39}  & 0.69      & \textbf{0.72$\pm$0.21} & 11.33$\pm$12.78          & 0.80    \\
\midrule
Swin  & Scratch          & N/A          & 0.54$\pm$0.31          & 11.73$\pm$11.46         & 0.58      & 0.70 $\pm$ 0.22        & 10.79$\pm$10.82          & 0.79    \\
Swin  & Self-pretraining & SMIT         & 0.63$\pm$0.23          & 8.95$\pm$8.57           & 0.67      & 0.74 $\pm$ 0.18        & 10.71$\pm$11.96          & 0.80    \\
Swin & Wild-pretraining & SMIT & \textbf{0.69$\pm$0.18} & \textbf{7.58$\pm$7.12} & \textbf{0.76} & \textbf{0.75 $\pm$ 0.15} & \textbf{9.69$\pm$8.97} & \textbf{0.85} \\
\bottomrule
\end{tabular}%
}
\vspace*{1ex}
\end{table}

\subsection{Tumor detection accuracy}
Both Swin and ViT models produced a higher tumor DR when using wild-pretraining compared to self-pretraining as well as scratch training. Swin showed the largest benefit with wild-pretraining (see Table~\ref{tab:tumor_seg_acc}). Wild-pretrained ViT showed larger improvement with respect to self-pretraining for the LRad than the LC dataset. The wild-pretrained CNN resulted in a slightly higher accuracy for LRad but lower DR for the LC dataset compared with its self-pretrained instance. The tumors detected by at least one model are used to report accuracy, yielding 139 for LRad dataset and 179 tumors for the LC dataset.  

\subsection{Segmentation accuracy}

Tumor segmentation accuracies for all three analyzed networks are shown in Table~\ref{tab:tumor_seg_acc}. As shown, both ViT and Swin models showed a higher accuracy with either pretraining approach compared to their scratch trained counterparts. Wild-pretrained Swin produced a significant accuracy improvement over self-pretrained model using both segmentation accuracy metrics (DSC p $<$ 0.0001, HD95 p $<$ 0.0001) on the LRad dataset but not significant on the LC data (DSC $p=$ 0.32, HD95 $p=$ 0.063). Wild-pretrained ViT resulted in slightly a higher DSC but similar HD95 as the self-pretrained model, indicating that either pretraining approach was similarly beneficial for ViT. Finally, the CNN model did not indicate any benefit from either pretraining approach compared to scratch trained CNN, indicating neither pretraining strategy was helpful in increasing tumor segmentation accuracy for the CNN model. 
\\

\subsection{Dependence of segmentation accuracy on tumor volumes}
\begin{figure}[t]
\vspace*{1ex}

    \begin{center}
    \includegraphics[width=1.0\columnwidth,scale=1.0]{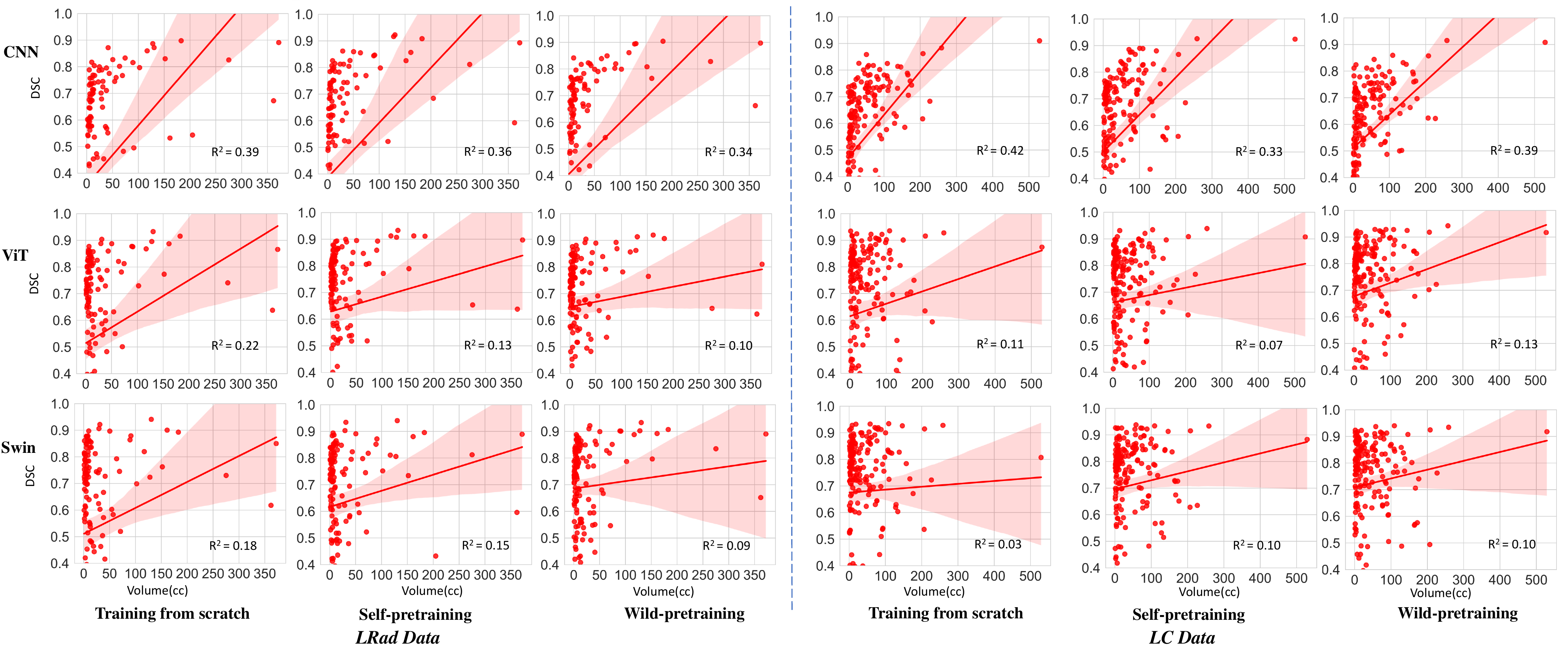}
       \caption{\small The scatter plot of DSC versus tumor volume (cc) to assess dependency of accuracy on the tumor volume for the analyzed architectures subjected to scratch, wild-pretraining and self-pretraining followed by fine-tuning for the two testing datasets. } 
       \label{fig:acc_vs_vol}
		\end{center}
\vspace*{1ex}
\end{figure}

Figure~\ref{fig:acc_vs_vol} shows the relationship between the DSC segmentation accuracy and tumor volumes for the LRad and LC datasets. In both datasets, ViT and Swin models showed a smaller dependence of segmentation accuracy to the tumor volume compare to the CNN model. Furthermore, wild-pretraining reduced the dependence of accuracy with respect to volume for the transformer models, indicating that wild-pretraining enabled the models to produce volume robust segmentations. 

\subsection{Robustness to CT imaging variations}
Table~\ref{tab:tumor_seg_contrast} shows the accuracy differences due to the presence and absence of CT contrast. All three models were significantly more accurate than their scratch-trained instances for contrast enhanced CTs. Only wild-pretrained ViT and Swin models showed a significant accuracy gain over scratch-trained models for non-contrast CT cases. On the other hand, wild-pretrained Swin was significantly more accurate than it's self-pretrained instance for both contrast and non-contrast CT cases. There was no statistical difference between wild- and self-pretrained CNN and ViT models for the non-contrast CT cases. 

\begin{table}[t]
\vspace*{1ex}

\centering
\caption{Tumor segmentation accuracy differences due to CT contrast evaluated on LRad dataset. Significance testing measured DSC differences between scratch, self- and wild-pretrained instances of the same architecture.}
\label{tab:tumor_seg_contrast}
\def\arraystretch{1.25}
    \resizebox{0.85\textwidth}{!}{
    \large
    \begin{tabular}{ll|ll|ll}
       Model & Training Strategy & Contrast (N=85) & p-value & Non-Contrast (N=54) & p-value \\
        \shline
         \textcolor{black}{CNN}  & \textcolor{black}{Scratch} & \textcolor{black}{0.47$\pm$0.33} & \textcolor{black}{0.014} & \textcolor{black}{0.35$\pm$0.33} & \textcolor{black}{0.28} \\
        \textcolor{black}{CNN}  & \textcolor{black}{Self-pretraining}  & \textcolor{black}{ 0.51$\pm$0.32}  & \textcolor{black}{0.72}& \textcolor{black}{0.36$\pm$0.32}  & \textcolor{black}{0.43}\\
        \textcolor{black}{CNN}  & \textcolor{black}{Wild-pretraining} & \textcolor{black}{\textbf{0.52$\pm$0.32}} & - & \textcolor{black}{\textbf{0.38$\pm$0.33}}  & - \\
        \midrule
        ViT  & Scratch & 0.60$\pm$0.30 & 0.00004 & 0.47$\pm$0.31 & 9.13e-6 \\
        ViT  & Self-pretraining  &  0.65$\pm$0.27  & 0.064& \textbf{0.65$\pm$0.19}  & 0.61\\
        ViT  & Wild-pretraining & \textbf{0.67$\pm$0.23} & - & 0.64$\pm$0.20  & - \\
        \midrule
        Swin  & Scratch  & 0.58$\pm$0.30 & 3.26e-6& 0.48$\pm$0.31 & 5.21e-7\\
        Swin  & Self-pretraining  & 0.66$\pm$0.23  & 0.0041& 0.60$\pm$0.23 & 6e-4\\
        Swin  & Wild-pretraining  & \textbf{0.70$\pm$0.19} & - & \textbf{0.68$\pm$0.16} & - \\
        \bottomrule
    \end{tabular}
    }
\vspace*{1ex}

\end{table}

\begin{figure}[t]
\vspace*{1ex}

    \begin{center}
	\includegraphics[width=0.99\columnwidth,scale=0.7]{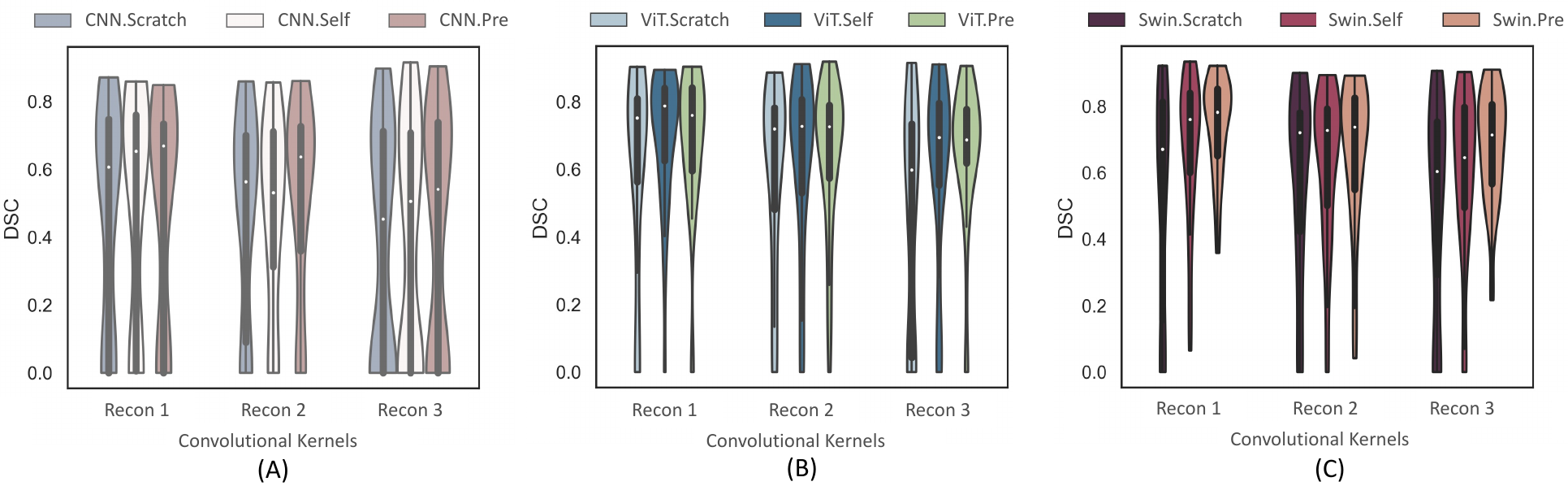}
		\vspace{0.00cm}\setlength{\belowcaptionskip}{-0.0cm}\setlength{\abovecaptionskip}{0.00cm}\caption{\small Influence of CT reconstruction kernel on segmentation accuracy with (A) CNN backbone and (B) ViT backbone, (c) Swin backbone. } \label{fig:boxplotKernels}
	\end{center}
\vspace*{1ex}
\end{figure}

Figure~\ref{fig:boxplotKernels} shows the impact of CT reconstruction convolution kernels on tumor segmentation accuracy using the public LRad dataset (n=139 CTs). The wild-pretrained Swin was significantly more accurate than self-pretrained Swin (p $<$ 0.001) with Recon 1 kernel and scratch trained Swin with Recon 1 (p $<$ 0.001) and Recon 3 kernels (p $=$ 0.006). On the other hand, accuracies were similar for wild-pretrained, self-pretrained, and scratch trained ViT and CNN models for the different reconstruction kernels. Both wild-pretrained ViT (p $=$ 0.008) and self-pretrained ViT were more accurate than scratch trained ViT (p $=$ 0.02) for Recon 3 kernels. 
\\
Next, we compared accuracy differences for the same model across the convolution kernels. Whereas, wild-pretraining resulted in similar accuracies across reconstruction kernels for Swin, the same architecture using self-pretraining produced less accurate segmentations with Recon 3 kernel compared to other two kernels. ViT showed similar mean accuracy for the three reconstruction kernels with self- and wild-pretraining. Scratch training showed a larger variation in the accuracy across the kernel reconstructions for both Swin and ViT architectures. The CNN network showed a wide variation in accuracy for all three kernels and training strategies.  

\begin{table}[t]
\vspace*{1ex}

\centering
\caption{Robustness of tumor segmentation to different scan reconstructions. Significance tests compared wild-pretrained to self-and scratch trained instances of the same architecture.}
\label{tab:tumor_seg_kernel}
\def\arraystretch{1.25}
    \resizebox{0.85\textwidth}{!}{
    \large
    \begin{tabular}{ll|lll|lll}
        \multirow{2}{*}{Model} &\multirow{2}{*}{Training Strategy} & Slice 2.5 mm & & & Slice 5 mm & & \\
        & & Recon 3 & Recon 1 & p-value & Recon 3 & Recon 1 & p-value \\ 
        \shline
         CNN  & Scratch  & 0.20$\pm$0.20 & 0.22$\pm$0.21 & 0.61 & 0.27$\pm$0.24 & 0.28$\pm$0.26&0.26 \\
        CNN  & Self-pretraining   & 0.21$\pm$0.20 & 0.22$\pm$0.20  &0.57 & 0.27$\pm$0.19 &  0.31$\pm$0.27 & 0.16\\
        CNN  & Wild-pretraining  &0.23$\pm$0.21 & 0.24$\pm$0.23 & 0.68& 0.30 $\pm$  0.25  & 0.34$\pm$0.31 &0.13 \\
        \midrule 
         ViT  & Scratch  & 0.47$\pm$0.34 & 0.54$\pm$0.32 & 0.08 & 0.49$\pm$0.34 &  0.50$\pm$0.30 & 0.64 \\
        ViT  & Self-pretraining   & 0.64$\pm$0.18 & 0.56$\pm$0.25  &0.019 & 0.58$\pm$0.26 &  0.51$\pm$0.23 & 0.14\\
        ViT  & Wild-pretraining  &0.67$\pm$0.16 & 0.58$\pm$0.26 & 0.077& 0.62$\pm$0.24 & 0.56$\pm$0.22 &0.11 \\
        \midrule
        Swin  & Scratch  & 0.52$\pm$0.32 &0.36$\pm$0.36 & 0.37&0.57$\pm$0.48  &0.47$\pm$0.34 & 0.12 \\
        Swin  & Self-pretraining  & 0.58$\pm$0.27 & 0.54$\pm$0.31 & 0.13& 0.52$\pm$0.28  & 0.49$\pm$0.30 & 0.058 \\
        Swin  & Wild-pretraining   & 0.70$\pm$0.18 & 0.66$\pm$0.21 &0.058 &0.62$\pm$0.28  & 0.58$\pm$0.26 & 0.036 \\
        \bottomrule
    \end{tabular}
    }
\vspace*{1ex}
\end{table}
Subset analysis of an additional set of 20 patients with paired reconstructions showed that wild-pretrained Swin produced significantly different accuracies between the two kernels when using 5 mm slices (p $=$ 0.036) but not with 2.5 mm slices (Table \ref{tab:tumor_seg_kernel}). On the other hand, self-pretrained Swin model showed similar but lower accuracy than the wild-pretrained Swin model for both kernel reconstructions and slice thicknesses. Wild-pretrained ViT and CNN models were robust to convolutional kernel differences with both slice thicknesses, as was the self-pretrained CNN model. However, both model architectures were less accurate than wild-pretrained Swin. The self-pretrained ViT showed significant accuracy difference between the convolutional kernels for the 2.5 mm slices but not the 5mm slices. These results indicate that slice thickness may impact the accuracy and robustness of the transformer models with respect to the convolutional kernels.

\begin{figure}[t]
\vspace*{1ex}
\centering

    \includegraphics[width=1.000\columnwidth,scale=0.7]{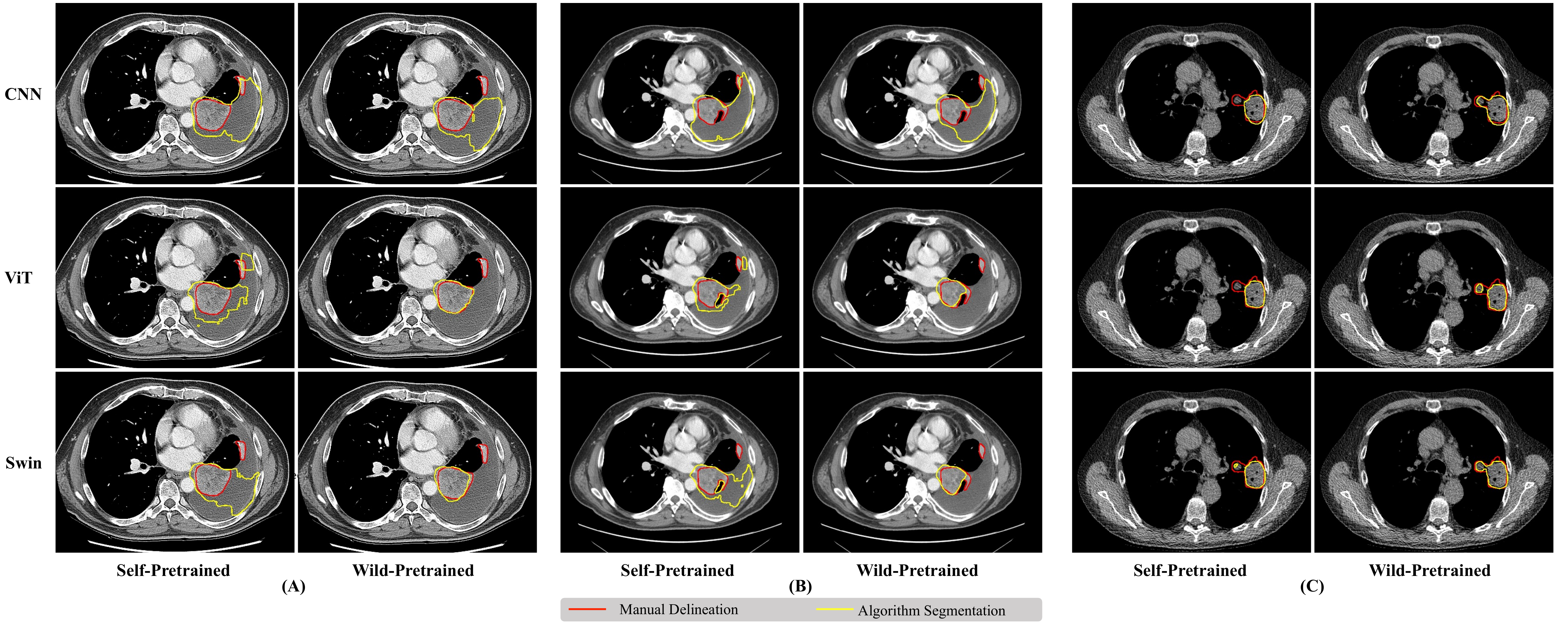}
    \caption{\small Segmentation (yellow contour) produced using CNN, ViT and Swin-based SMIT model using wild-pretaining and self-pretraining. (A) Contrast CT scan using GE lung reconstruction (B)Contrast CT scan using GE standard reconstruction (C) Non-contrast CT scan using GE lung reconstruction. Manual delineation is shown in red and algorithm delineation is shown in yellow.} 
    \label{fig:seg_show_supp}
\vspace*{1ex}
\end{figure}

Figure~\ref{fig:seg_show_supp} shows the segmentations produced by the three networks when subjected to self- and wild-pretraining on two representative patients with contrast CT scan using (A) Lung reconstruction kernel, (B) Standard reconstruction kernel and (C) non-contrast CT scan. In the case of both ViT and Swin models, wild-pretrained instances better approximate the manual delineations irrespective of image contrast compared to the self-pretrained instances. The results are similar for both self- and wild-pretrained instances when applied to the CNN model. 

\begin{table}[!h]
\vspace*{1ex}

\centering
\caption{Model performance on the phantom image with 8 lesions.}

\label{tab:phantomresults}
\def\arraystretch{1.25}
\resizebox{0.65\textwidth}{!}{
\begin{tabular}{llllll}
Model& Training Strategy & DSC & HD$_{95}$ & DR  \\ \shline

CNN & Self-pretraining & 0.71 $\pm$ 0.04 & 4.00 $\pm$ 1.13 & 0.25 \\ 
CNN & Wild-pretraining & 0.00 $\pm$ 0.00 & $ NA$ & 0.00 \\
\midrule

ViT & Self-pretraining & 0.81 $\pm$ 0.06 & 2.60 $\pm$ 0.49 & 0.625 \\ 
ViT & Wild-pretraining & 0.81 $\pm$ 0.05 & 3.24 $\pm$ 2.01 & 0.75 \\
\midrule

Swin & Self-pretraining  & 0.86 $\pm$ 0.01 & 1.46 $\pm$ 0.06 & 0.5 \\ 
Swin& Wild-pretraining & 0.86 $\pm$ 0.06 & 1.75 $\pm$ 0.74 & 0.875 \\
\bottomrule
\end{tabular}
}
\label{tab:model_performance}
\vspace*{1ex}

\end{table}

\begin{figure}[t]
\vspace*{1ex}

    \centering
    \includegraphics[width=0.9\columnwidth,scale=0.7]{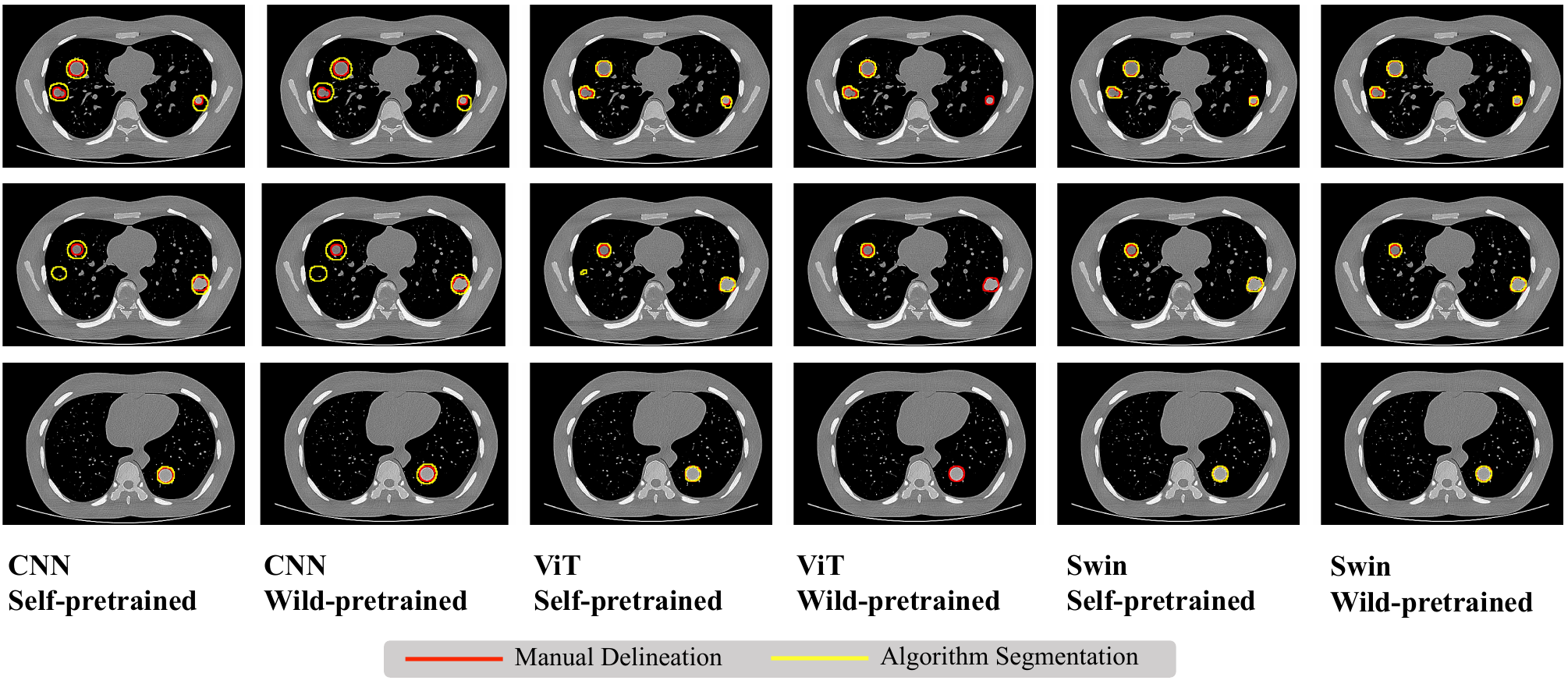}
    \caption{Results on the phantom image scan (dataset). Yellow contour indicates the model segmentation and red contour indicates the phantom ground truth.}
    \label{fig:results_phantomlabel}
\vspace*{1ex}
\end{figure}

Table~\ref{tab:phantomresults} shows the detection rate (DR) and the segmentation accuracy for the detected tumors on the lung phantom dataset. The wild-pretrained Swin transformer achieved the highest DR across all models. The self-pretrained Swin produced a similar DSC accuracy as the wild-pretrained Swin, and a relatively better HD95 metric, but a lower detection rate indicating difficulty in detecting all lesions. Similar scores were observed for ViTs, where the detection rate was higher and the HD95 lower, suggesting a potential trade-off in finer detection abilities with the self-pretrained models. The CNN model on the other hand showed very low tumor detection capability compared to the transformer models. Figure \ref{fig:results_phantomlabel} shows the segmentation performance of self-pretrained and wild-pretrained models using CNN, ViT and Swin on the phantom dataset. 
\\

\subsection{Ablation analysis}
Ablation analysis was performed using only the Swin backbone as this network emerged as the most accurate compared to the other two networks.

\subsubsection{Impact of different pretext tasks}
Table \ref{tab:ablation_pretextTasks_seg_acc} shows the detection rate and segmentation accuracy for the detected tumors for the network subjected to wild- and self-pretraining using multiple pretext tasks. As shown, wild-pretraining with the SMIT pretext tasks that combined MIP, ITD, and MPD losses resulted in the highest tumor segmentation accuracy. Also, wild-pretraining with SMIT task was more accurate than self-pretraining using the same pretext task. MIP pretext task resulted in higher accuracy compared to other individual pretext tasks including contrastive loss, indicating that the generative image prediction task, which forces the network to learn the local and global structure was really important for learning good feature representation. In particular, self-pretraining with MIP resulted in the highest detection rate. All other pretext tasks resulted in similar accuracies for the self- and wild-pretrained versions using those same tasks. Contrastive pretext task resulted in among the worse accuracy, indicating that contrastive loss that encourages the network to learn features to distinguish images as a whole is less beneficial for segmenting individual structures such as tumors. 

\subsubsection{Impact of two stage pretraining} 
As shown in Table \ref{tab:ablation_pretextTasks_seg_acc}, two-stage pretraining improved tumor detection rate compared to self-pretraining but was less accurate than the model that only used wild-pretraining. The trend was also reflected in tumor segmentation accuracy, indicating that the two-step pretraining strategy is not beneficial for accuracy.

\subsubsection{Impact of fine-tuning with homogeneous dataset}
We evaluated the impact of pretraining strategies when the network was subjected to fine-tuning with a homogeneous dataset, wherein fine-tuning was performed with a subset of 279 non-contrast CT scans acquired at 3mm slice thickness using the Siemens scanner. Testing was performed on the LRad dataset that is different from the fine-tuning dataset. As shown in Table ~\ref{tab:tumor_seg_contrast_w_only_Siemens}, for both ViT and Swin networks, wild-pretrained models resulted in higher accuracy for both contrast and non-contrast CTs compared to the self-pretrained models, indicating improved generalization capability with wild-pretrained models. 

\begin{table}[t]
\vspace*{1ex}

\centering
\caption{Impact of pretext tasks on down-stream accuracy applied to Swin model.}
\label{tab:ablation_pretextTasks_seg_acc}
\def\arraystretch{1.25}
\resizebox{0.85\textwidth}{!}{%
\begin{tabular}{ll|lll|lll}
\multirow{2}{*}{\begin{tabular}[c]{@{}l@{}}Pretext\\ Task\end{tabular}} & \multirow{2}{*}{Training Strategy} & LRad &  &  & LC &  &  \\
 &  & DSC & HD$_{95}$ & DR & DSC & HD$_{95}$ & DR \\ \shline
MIP & Self-pretraining & 0.67 $\pm$ 0.19 & 8.71$\pm$8.26 & \textbf{0.78} & 0.72$\pm$0.18 & 11.81$\pm$15.52 & \textbf{0.85} \\
MIP & Wild-pretraining & 0.64$\pm$0.25 & 8.58$\pm$8.77 & 0.72 & 0.73$\pm$0.17 & 11.01$\pm$11.92 & 0.84 \\ \midrule
ITD & Self-pretraining & 0.64$\pm$0.24 & 8.62 $\pm$ 8.35 & 0.72 & 0.71 $\pm$ 0.21 & 11.11$\pm$12.14 & 0.81 \\
ITD & Wild-pretraining & 0.64$\pm$0.24 & 8.33$\pm$7.93 & 0.69 & 0.72$\pm$0.20 & 10.61$\pm$12.33 & 0.82 \\ \midrule
ITD \& MPD & Self-pretraining & 0.66 $\pm$ 0.22 & 8.62$\pm$8.36 & 0.72 & 0.71$\pm$0.21 & 11.56$\pm$ 14.24 & 0.83 \\
ITD \& MPD & Wild-pretraining & 0.66$\pm0.21$ & 8.62$\pm$7.22 & 0.72 & 0.72$\pm$0.20 & 10.46$\pm$9.68 & 0.82 \\ \midrule
Contrastive & Self-pretraining & 0.63$\pm$0.24 & 9.29 $\pm$ 9.23 & 0.69 & 0.71 $\pm$ 0.23 & 11.44$\pm$ 12.93 & 0.79 \\
Contrastive & Wild-pretraining & 0.64$\pm$0.24 & 9.08$\pm$9.68 & 0.73 & 0.71$\pm$0.21 & 11.34$\pm$12.05 & 0.79 \\ \midrule
SMIT & Self-pretraining & 0.63$\pm$0.23 & 8.95$\pm$8.57 & 0.67 & 0.74 $\pm$ 0.18 & 10.71$\pm$11.96 & 0.80 \\
SMIT & Wild-pretraining & \textbf{0.69$\pm$0.18} & \textbf{7.58$\pm$7.12} & 0.76 & \textbf{0.75 $\pm$ 0.15} & \textbf{9.69$\pm$8.97} & 0.85 \\
SMIT & Wild- + Self-pretraining & 0.65$\pm$0.21 & 8.52$\pm$7.72 & 0.72 & 0.73$\pm$0.17 & 10.65$\pm$11.41 & 0.82 \\
\bottomrule
\vspace*{1ex}
\end{tabular}%
}
\end{table}

\begin{table}[t]
\vspace*{1ex}

    \centering
    \caption{Testing accuracy of models fine-tuned with with homogenous task data (contrast enchanced CT, Siemens, Recon 1).}
    \label{tab:tumor_seg_contrast_w_only_Siemens}
    \def\arraystretch{1.25}
    \large
    \resizebox{0.8\textwidth}{!}{
        \begin{tabular}{ll|ll|ll}
        Model & Training Strategy & Contrast (N=85) & p-value & Non-Contrast (N=54) & p-value \\
        \shline
        ViT  & Self-pretraining  & 0.61$\pm$0.27  & 0.0869  & 0.59$\pm$0.24 & 0.440\\ 
        ViT  & Wild-pretraining  & 0.63$\pm$0.25  &  & 0.56$\pm$0.25  & \\
        \midrule
        Swin  & Self-pretraining   & 0.65$\pm$0.24   & 0.019  & 0.63$\pm$0.20&0.412\\
        Swin  & Wild-pretraining   &  0.69$\pm$0.20  & & 0.65$\pm$0.19   \\
        \bottomrule
        \end{tabular}
    }
\vspace*{1ex}

\end{table}

\begin{figure}[t]
\vspace*{1ex}
    \begin{center}
    \includegraphics[width=0.85\columnwidth,scale=0.5]{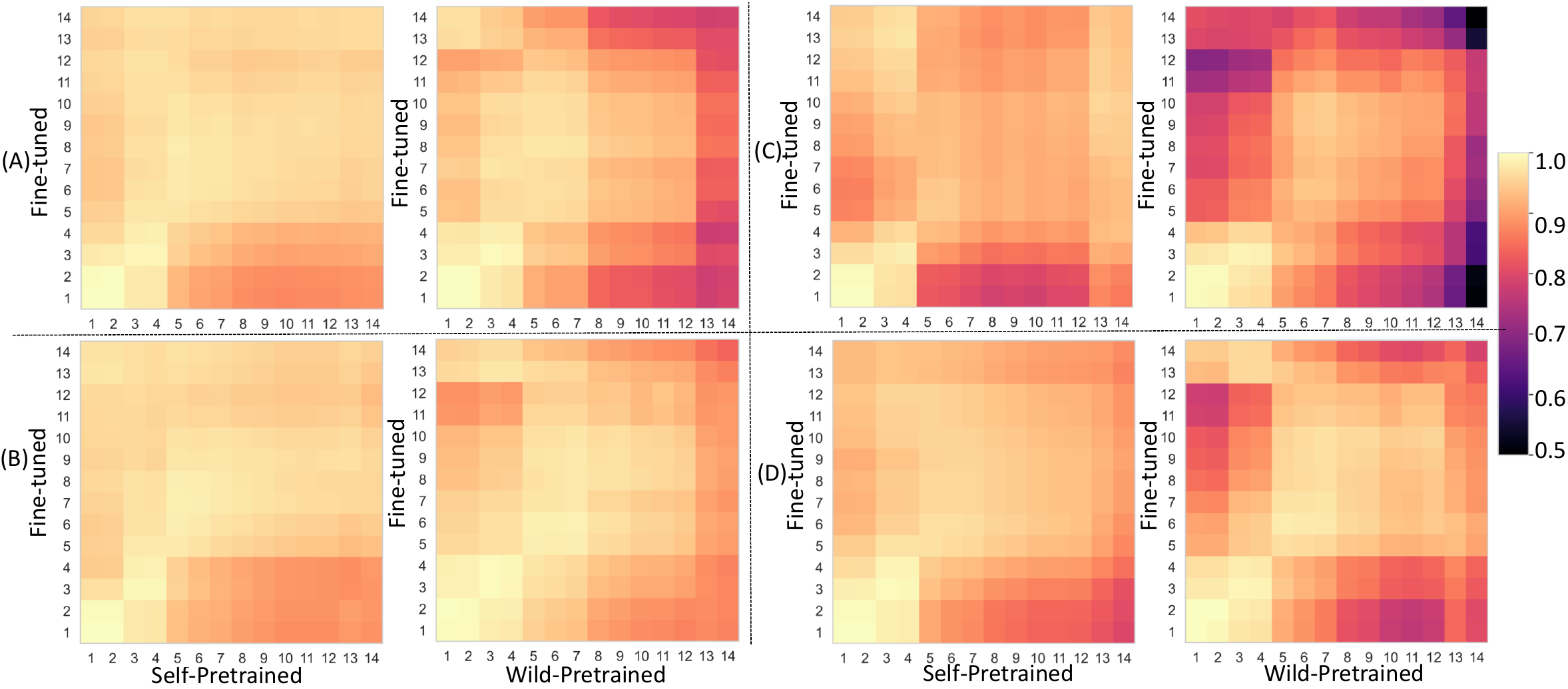}
		\vspace{-0.00cm}\setlength{\belowcaptionskip}{-0.0cm}\setlength{\abovecaptionskip}{0.00cm}\caption{\small CKA analysis to measure feature reuse for (A) contrast and (B) non-contrast CT as well as CT images reconstructed using (C) Recon 1 and (D) Recon 3. } 
    \label{fig:cka_contrast}
    \end{center}
\vspace*{1ex}
\end{figure}

\begin{figure}[t]
\vspace*{1ex}
    \begin{center}
    \includegraphics[width=1.00\columnwidth,scale=0.7]{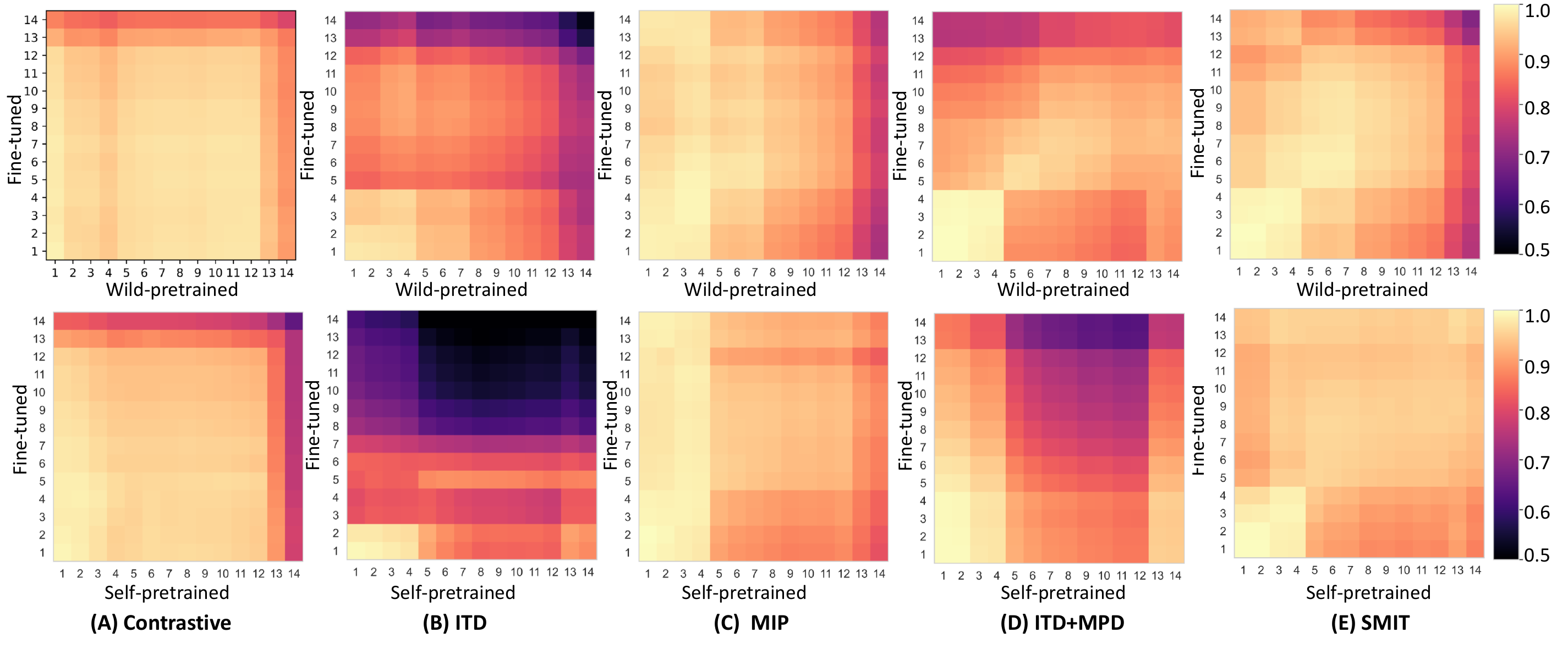}
    \vspace{-0.00cm}\setlength{\belowcaptionskip}{-0.0cm}\setlength{\abovecaptionskip}{0.00cm}\caption{\small CKA analysis performed on the Swin backbone using different pretext tasks including (A) Contrastive, (B)ITD, (C)MIP, (D)ITD+MPD and (E) SMIT with wild- and self-pretraining.} 
    \label{fig:cka_diff_method}
    \end{center}
\vspace*{1ex}
\end{figure}

\subsection{Feature reuse analysis}

\subsubsection{Feature reuse for different CT acquisitions}
Figure~\ref{fig:cka_contrast} shows the feature reuse using CKA analysis between different pretrained and fine-tuned models for two different CT reconstruction convolution kernels using contrast and non-contrast as well as Recon 1 and Recon 3 from the public LRad dataset. These two reconstructions were evaluated due to larger and similar prevalence of the cases between these reconstructions. 

There was high feature reuse for the self-pretrained model for both contrast and non-contrast CTs (Figure~\ref{fig:cka_contrast}(A, B)). On the other hand, wild-pretrained model resulted in resulted in larger deviations of features close to the later encoder layer (13 and 14) for contrast compared to non-contrast CTs (Figure~\ref{fig:cka_contrast} A and B). 
Analysis of Recon 1 and Recon 3 showed higher feature reuse with wild-pretrained models for the lower levels compared to later layer (13 and 14) as shown in Figure~\ref{fig:cka_contrast} C and D.  

\subsubsection{Impact of pretext tasks on feature reuse}
Figure~\ref{fig:cka_diff_method} shows the feature reuse for the Swin network using CKA when the network was subjected to self- and wild-pretraining with the different pretext tasks. Analysis was performed on the LRad testing data. 

There was a considerable variation in the reuse of the features for the same network architecture (Swin) when wild-pretrained with different pretext tasks as shown in Figure~\ref{fig:cka_diff_method} (A) to (E). Concretely, contrastive task resulted in the highest feature reuse even across different feature layers (off-diagnoal entries in the CKA matrix). ITD, a global image feature matching loss, resulted in the lowest feature reuse, followed by ITD+MPD, MIP, and SMIT (Figure~\ref{fig:cka_diff_method}). SMIT, which uses a combination of ITD, MIP and MPD, the latter two are spatial locality context losses, resulted in higher feature reuse in the lower (1 to 4) and middle level features (5 to 9) compared to higher level (10 to 14) layer features. In addition, the features across different layers (off-diagonal entries of the CKA matrix) were different between wild-pretrained and fine-tuned features for SMIT, MPD, and ITD tasks but not for the contrastive learning task. Self-pretraining (Figure~\ref{fig:cka_diff_method} E) with SMIT resulted in more differentiation of off-diagonal features (layers 5 to 14 compared to lower level features 1 to 4) but such differentiation was to a lesser degree than wild-pretrained SMIT. In general, wild-pretraining resulted in feature changes especially close to the later stage encoder layers (13 and 14) for all the pretext tasks when compared to self-pretraining. 

For both self-pretrained and wild-pretrained models, a high similarity in the lower level (layer 1-4) is present regardless of the pretext tasks, except for the ITD task for the self-pretrained model. On the other hand, across all tasks, feature dissimilarity increases close to output layer as the task varies between wild-pretraining and fine-tuning. This separation also leads to improved accuracy of the wild-pretrained Swin network.

\section{Discussion}
This study performed a comprehensive analysis of robustness to CT imaging variations due to scratch versus self- versus wild- pretraining of two transformer and one convolutional neural network architectures. All three network architectures showed improved accuracy and robustness when using any pretraining strategy compared to scratch training, indicating that pretraining provides useful features for downstream tasks. Importantly, Swin network showed a significant accuracy improvement with wild-pretraining compared to self-pretraining for contrast and non-contrast CTs as well as for Recon 1 and Recon 3 convolution kernels. On the other hand, wild-pretrained ViT showed a trend towards higher accuracy but not a significant accuracy improvement over self-pretrained model. The CNN-based network showed similar accuracies for both pretraining strategies, indicating that self-pretraining is sufficient for the CNN network. Of the three architectures, wild-pretrained Swin network was the most accurate for all analyzed scenarios including the lung CT phantom case. 

Our results are consistent with findings from natural images that demonstrated improvements in accuracy and fine-tuning efficiency with pretraining~\cite{Goyal2021}. We also found that Swin models benefit more from wild-pretraining compared to ViT. Relatedly, analysis with pretext tasks indicated that combination of masked image prediction with global and patch token distillation that combines global and local contextual similarities improves testing accuracy more than the individual losses and also leads to higher feature reuse in the lower layers. In particular, the use of masked image prediction, forces the networks to learn the spatial inductive bias during pretraining, which in turn increases feature reuse in the lower level layers. On the other hand, contrastive pretext task, which only requires the model to learn the embedding for the image as a whole was not effective in increasing detection or segmentation accuracy with either pretraining methods. These results also suggest that fine-tuning of pretrained models can be optimized close to higher level features to further increase fine-tuning efficiency. 

Feature reuse analysis performed to understand why the networks differ in accuracy showed a diverse feature reuse pattern between the self- and wild-pretrained Swin model, which also varied with the choice of pretext task used for pretraining. In particular, wild-pretraining resulted in more feature reuse at the lower level features across all pretext tasks but showed a specialization of features with fine tuning close to the output layer, which is important for higher accuracy on the task. Self-pretraining, on the other hand, resulted in similar patterns of feature reuse at both low and high level layers, thus resulting in a lower accuracy in the analyzed task. 

Finally, analysis of contrast, non-contrast, and different kernel reconstructions showed a consistent pattern, wherein wild-pretraining resulted in a higher feature differentiation close to the output layer compared to the self-pretrained models, which ultimately resulted in higher accuracy with wild-pretraining. 

 \section{Conclusions}
Both wild-pretrained and self-pretrained models enhance lung tumor segmentation across various architectures, including CNN, ViT and Swin. Wild-pretrained networks were more robust to analyzed CT imaging differences for lung tumor segmentation than self-pretrained methods. Swin architecture benefited from such pretraining more than ViT and CNN-based architecture.  

\section{Acknowledgement}
This research was partially supported by the NCI R01CA258821 and the Memorial Sloan Kettering Cancer Center Support Grant/Core Grant NCI P30CA008748.

\section*{References}
{\tiny
\bibliographystyle{IEEEtran}
}
\bibliography{midl-samplebibliography}


\end{document}